\documentclass[conference]{IEEEtran}
\IEEEoverridecommandlockouts
\usepackage{cite}
\usepackage{amsmath,amssymb,amsfonts}
\usepackage{algorithmic}
\usepackage{graphicx}
\usepackage{textcomp}
\usepackage{xcolor}
\usepackage{tikz}

\usepackage[export]{adjustbox}
\usepackage{cleveref}
\Crefformat{figure}{#2Fig.~#1#3}
\Crefmultiformat{figure}{Figs.~#2#1#3}{ and~#2#1#3}{, #2#1#3}{ and~#2#1#3}
\Crefformat{equation}{#2Eq.~#1#3}
\Crefmultiformat{equation}{Eqs.~#2#1#3}{ and~#2#1#3}{, #2#1#3}{ and~#2#1#3}
\usepackage{subcaption}
\usepackage{url}

\usepackage{xcolor}
\usepackage{multirow}
\usepackage{tabularx}
\usepackage{booktabs}

\def\BibTeX{{\rm B\kern-.05em{\sc i\kern-.025em b}\kern-.08em
    T\kern-.1667em\lower.7ex\hbox{E}\kern-.125emX}}

\begin{document}
\IEEEoverridecommandlockouts
% \IEEEpubid{\makebox[\columnwidth]{978-1-4799-7492-4/15/\$31.00~
% \copyright2015
% IEEE \hfill} \hspace{\columnsep}\makebox[\columnwidth]{ }} 
\title{Estimating Musical Surprisal in Audio

%  GW: Modeling Musical Surprisal in Audio \\ Using a Transformer Model

% GW: A Learning-based Model of Musical Surprisal in Audio

% GW: Modeling Musical Surprisal in Audio \\ via Latent Representations and Prediction 

\thanks{ 
Supported by the European Research Council ERC under grants 101019375 (\textit{“Whither Music?”}) [MRB, GW] and 787836 (\textit{NEUME}) [GC]; FrontCog grant ANR-17-EURE-0017 [GC]; and by a joint research project between SONY CSL and JKU [MRB, SL, GW].}
}

\author{
\IEEEauthorblockN{Mathias Rose Bjare\IEEEauthorrefmark{1}, Giorgia Cantisani\IEEEauthorrefmark{2}, Stefan Lattner\IEEEauthorrefmark{3} and Gerhard Widmer\IEEEauthorrefmark{1}\IEEEauthorrefmark{4}}
\IEEEauthorblockA{\IEEEauthorrefmark{1}Institute of Computational Perception\\ Johannes Kepler University (JKU) Linz, Austria
% Georgia Institute of Technology, Atlanta, Georgia 30
% 332--0250\\
}
\IEEEauthorblockA{\IEEEauthorrefmark{2}
Laboratoire des Systèmes Perceptifs \\
ENS, PSL University, CNRS, Paris, France\\
}

\IEEEauthorblockA{\IEEEauthorrefmark{3}Sony Computer Science Laboratories (CSL), Paris, France}
\IEEEauthorblockA{\IEEEauthorrefmark{4}LIT AI Lab, Linz Institute of Technology, Linz, Austria\\
Email: \url{mathias.bjare@jku.at}
}
% \IEEEauthorblockN{1\textsuperscript{st} Given Name Surname}
% \and
% \and
% \and
% \and
% \and
}
% \IEEEpubid{©20XX IEEE.  Personal use of this material is permitted. Permission from IEEE must be obtained for all other uses, in any current or future media, including reprinting/republishing
% this material for advertising or promotional purposes, creating new collective works, for resale or redistribution to servers or lists, or reuse of any copyrighted component of this work in other works.}
% \IEEEpubid{\makebox[\columnwidth]{“\copyright 20XX IEEE.  Personal use of this material is permitted. Permission from IEEE must be obtained for all other uses,
% in any current or future media, including reprinting/republishing this material for advertising or promotional purposes, creating new collective works, for resale or redistribution to servers or lists, or reuse of any copyrighted component of this work in other works.” \hfill} \hspace{\columnsep}} 

\maketitle
\begin{abstract}
%In the context of modeling the surprisal of musical content using computational methods, it has been proposed for symbolic music to use the information content (IC) of one-step musical event predictions of an autoregressive critic model as a proxy for musical surprisal. 
In modeling musical surprisal expectancy with computational methods, it has been proposed to use the information content (IC) of one-step predictions from an autoregressive model as a proxy for surprisal in symbolic music.
With an appropriately chosen model, the IC of musical events has been shown to correlate with human perception of surprise and complexity aspects, including tonal and rhythmic complexity. This work investigates whether an analogous methodology can be applied to music audio. We train an autoregressive Transformer model to predict compressed latent audio representations of a pretrained autoencoder network. We verify learning effects by estimating the decrease in IC with repetitions. We investigate the mean IC of musical segment types (e.g., A or B) and find that segment types appearing later in a piece have a higher IC than earlier ones on average.
We investigate the IC's relation to audio and musical features and find it correlated with timbral variations and loudness and, to a lesser extent, dissonance, rhythmic complexity, and onset density related to audio and musical features.
Finally, we investigate if the IC can predict EEG responses to songs and thus model humans' surprisal in music. We provide code for our method on \url{github.com/sonycslparis/audioic}.

\end{abstract}
%TC:ignore
% \mb{After consulting with a frequent ICASSP submitter, it seems papers typically invent their own Index Terms, although the template mentions using ``valid keywords'', which I assume are the ones from https://www.ieee.org/content/dam/ieee-org/ieee/web/org/pubs/ieee-thesaurus.pdf, from which the following seem related: Computer applications Music, Music, Music information retrieval, Computer generated music, Rhythm, Timbre, Information Theory, Neural networks, Generative AI. }
%
\begin{tikzpicture}[remember picture, overlay]
  \node[anchor=south, yshift=10mm, text width=2\linewidth, align=center, 
        draw, font=\small] at (current page.south) {
    © 2025 IEEE. Personal use of this material is permitted. Permission from IEEE must be obtained for all other uses, in any current or future media, including reprinting/republishing this material for advertising or promotional purposes, creating new collective works, for resale or redistribution to servers or lists, or reuse of any copyrighted component of this work in other works.
  };
\end{tikzpicture}
\section{Introduction}
\label{sec:intro}
Surprisal, as estimated by the \textit{information content} (IC) or negative log-likelihood of an autoregressive sequence model, has been proposed as a proxy for estimating perceived musical surprise as experienced by human listeners \cite{meyer,idyom_conklin,idyom,musexp}. With an appropriately chosen model, the IC of musical events has been shown to correlate with human perception of surprise and with complexity aspects, including tonal and rhythmic complexity \cite{complic,BjareLW23}. The analysis of music with IC offers a way to quantify information-theoretic hypotheses about music and music perception \cite{gold2019predictability}
%(e.g., uniform information density \cite{gold2019predictability}) 
and has been shown to work as an interesting conditioning signal for generative models \cite{wang2014guided,collins2016developing, bjare2024controlling}. 
With a few exceptions \cite{dubnov2007audio,masclef2023deep,abrams2022retrieving}, most work on computing surprisal in music data has been limited to the discrete symbolic domain (e.g., sheet-music or MIDI), for which music can be serialized as condensed, interpretable sequences of musical events where the surprisal can be identified with specific musical events, such as an unexpected pitch breaking with the expectation of an established tonality, or sudden change of rhythm. 
The condensed discrete nature of symbolic encodings makes them suitable for studying suprisal with language models \cite{BjareLW22,bjarelattner,BjareLW23,bjare2024controlling} like Transformers\cite{VaswaniSPUJGKP17}. However, as opposed to audio representations, symbolic music has limited availability and fails to capture all musical aspects, including acoustic properties like timbre and dynamics, that are critical to many musical genres, including popular music, and might influence the experience of surprisal. Therefore, estimating surprisal using audio would be more general than using symbolic data.

Analyzing surprisal in full-length music audio has until recently \cite{evans2024long} been impractical with invertible audio representations that faithfully preserve the audio content. This has been due to the low compression rates of such representations, causing the sequences to be restrictively long. As such, previous works have calculated IC using preselected audio features \cite{dubnov2007audio,abrams2022retrieving}, chosen using domain expertise of the studied music that potentially bias facets of the investigation.

% \IEEEpubidadjcol
In this work, we analyze surprisal aspects of full-length % multi-stem 
music recordings without using preselected audio features. 
More specifically, we train a GIVT model \cite{Tshannen2023givt} and estimate IC from its one-step predictions on audio sequences encoded into the invertible audio representations of Music2Latent \cite{pasini2024music2latent}. We use the mean IC (information density) of music segments to show the effect of learning by analyzing the reduction in IC on repeated segments. Furthermore, we find that segment types that appear later in pieces have higher IC, suggesting that they have been composed to contrast earlier segments musically. We correlate the IC with audio signal features and musical features typically associated with surprisal and complexity. Finally, using conventional cortical tracking methodologies for validating computational models of human expectations, we found that our IC significantly predicts EEG brain responses to songs, supporting our methodology’s potential for modeling human surprisal in music.

\section{Related work}
In the symbolic music domain, IC, as a proxy for musical surprisal, has most notably been studied with the variable order Markov-model IDyOM \cite{idyom_conklin}.
Several behavioral and neural studies already support IDyOM modeling of human melodic expectation \cite{idyom, pitch_per, di2020cortical, hansen2014predictive, bianco2020pupil, moldwin2017statistical}, and its predictions were shown to correlate with tonal and rhythmic complexity \cite{complic,BjareLW23}, typically associated with musical surprisal.
The model is, however, limited to monophonic symbolic music stimuli. In the work of \cite{bjare2024controlling}, the authors propose an IC-based technique for estimating surprisal in polyphonic symbolic music and show the IC to correlate with tonal and rhymic complexity using solo piano performances.
In the audio domain, suprisal has been calculated using human-selected audio features. The Audio Oracle \cite{dubnov2007audio} analyzes surprisal using \textit{information rate} calculated from self-similarities of audio features. The method was shown to identify high surprisal with segment boundaries.
In \cite{abrams2022retrieving}, IC of a D-REX \cite{Skerritt-DavisE18,skerritt2019model} model is related to the MEG brain response of human participants. IC is calculated using 15 different audio features and Bayesian inference.
Although not related to temporal suprisal, \cite{masclef2023deep} uses the KL-divergence of a diffusion model to approximate the likelihood of 5-second clips. This is used to reproduce the inverted U-shape relation between the total IC of music clips and listener preference presented in  \cite{gold2019predictability}. The model does not rely on audio features; however, it ignores causality and memory aspects of surprisal.
\section{Method}
\subsection{Audio Encoding}
Instead of modeling surprisal directly in the audio domain, we encode our data into the invertible 64-dimensional latent representations of Music2Latent \cite{pasini2024music2latent}, an autoencoder % \cite{KingmaW13}
with a continuous latent bottleneck, based on consistency models \cite{SongD0S23}. Thus, the audio signal is encoded in a 64-channel sequence of $\sim$11Hz. 
We choose an autoencoder representation over a contrastively learned representation \cite{BaevskiZMA20,SpijkervetB21} to ensure that the representation preserves the signal information. Furthermore, we chose Music2Latent over discrete representations \cite{jukebox,ZeghidourLOST22,DBLP:journals/tmlr/DefossezCSA23,KumarSLKK23}, due to its low frame rate allowing us to train autoregressive models on full-length music pieces.
\subsection{Modelling Suprisal}
When dealing with symbolic music or discrete audio representations, surprisal can be obtained directly from the (learned) next-step prediction probability mass function. In the context of neural networks, this means having a finite vocabulary of tokens and typically implies training a Transformer to do masked prediction \cite{VaswaniSPUJGKP17,radford2019language} with a softmax cross-entropy loss. More specifically, given a sequence of discrete tokens $x_{1,2,...,T}$ from a finite vocabulary, the IC of sequence element $x_{t}=i$ is given by %\stefan{necessary?} \mb{the reason for putting it here, was to emphasize the difference between discrete and continuous IC (the latter being somehow non-standard and have maybe even not been attempted before). If it is all crystal clear, we can shorten this section.} \gw{I would leave this in; I think this passage reads well; otherwise it would be too dense.}
\begin{equation}
    \text{IC}\left(x_{t} = i \vert x_{<t}\right) = -\log \left(\text{Softmax}\left(f_{\theta}\left(x_{<t}\right)\right)_{i}\right),
    \label{eq:ic_dis}
\end{equation} where $f_{\theta}$ is a Transformer model, outputting the softmax logits. As such, IC is bounded by the interval $[0, \infty[$. Working with continuous representations, the vocabulary is infinite, and we cannot estimate a probability mass function. Instead, we model the IC using the probability density of the next frame. Similarly to \cite{Tshannen2023givt}, we replace the softmax function in \eqref{eq:ic_dis} with a Gaussian mixture model (GMM) parameterized by the logits of the Transformer $f_{\theta}$, i.e.,
\begin{equation}
    \text{IC}\left(x_{t} = v \vert x_{<t}\right) = -\log \left(
        p_{\mathit{GMM}}\left(x_{t}=v ;f_{\theta}\left(x_{<t}\right)\right)
    \right), 
    \label{eq:ic_con}
\end{equation} where $p_{\mathit{GMM}}$ is the differentiable probability density function of a GMM with parameters $f_{\theta}(x_{<t})$ (Parameterization in \cref{sec:modeltrain}). We note that the IC of our model is unbounded. 

\section{Training Details}
% \begin{table*}[t]
% \centering
% \caption{}
% \label{tab:res}

% \begin{tabular}{lcccccccccccc}
% \toprule
%      & \multicolumn{2}{c}{Repetition} & \multicolumn{2}{c}{Segment} & \multicolumn{8}{c}{Complexity} \\
%  & $\rho$ & $p$-val & $\rho$ & $p$-val & $\rho_{di}$ & $p$-val & $\rho_{rh}$ & $p$-val & $\rho_{de}$ & $p$-val & $\rho_{lo}$ & $p$-val  \\
%  \cmidrule(r){2-3}
%  \cmidrule(r){4-5}
%  \cmidrule(r){6-13}
% %\midrule
% % no mean segment
% % MU & -0.06 & 4.7 $\times 10^{-2}$ & x & x & x & x & x & x  & & & &\\
% % ME & -0.11 & 3.6 $\times 10^{-2}$ & x & x & x & x & x & x & & & &\\
% % IN & -0.17 & 1.4 $\times 10^{-141}$& x & x & x & x & x & x & & & & \\
% MU & -0.76 & 4.3 $\times 10^{-4}$ & 0.89 & 6.8 $\times 10^{-3}$& 0.07 & 3.5 $\times 10^{-10}$ & 0.07 & 1.7 $\times 10^{-10}$ & -0.03 & 2.5 $\times 10^{-2}$ & 0.45 & 0.0 \\
% ME & -0.64 & 2.4 $\times 10^{-2}$ & 0.85 & 1.3 $\times 10^{-2}$ & 0.21 & 5.0 $\times 10^{-72}$  & 0.09 & 8.2 $\times 10^{-12}$ & 0.14 & 3.7 $\times 10^{-16}$ & 0.57 & 0.0 \\
% IN & -0.52 & 8.3 $\times 10^{-3}$ & 1 & 0 & 0.20 & 0 & 0.11 & 0 & 0.07 & 0 & 0.70 & 0 \\
% \bottomrule
% \end{tabular}
% \end{table*}
\subsection{Data}
For training our model, we use a proprietary multi-stem dataset (PR) consisting of 19,701 pieces of popular music. %\gw{(The multi-stem aspect will be useful for the experiment described in ... below.)}
The multi-stem aspect will be useful for our experiment on correlating the surprisal estimations with EEG responses to singing voices in \cref{sec:eeg} since it contains vocal stems.
The dataset is split into 17,730 and 1,969 pieces for training and evaluation, respectively. In addition, we evaluate our model on the multi-stem datasets of MedleyDB 2.0 (ME) \cite{bittner2016medleydb} and MUSDB18 (MU) \cite{musdb18}. For the former, we use the 69 pieces annotated with Pop, Rock, or that have genre annotation. For the latter, we use the 74 pieces tagged with Pop, Country, Reggae, and Rock that are absent in ME. For all datasets, we preprocess the audio by averaging the stems, converting to mono, downsampling to 22050Hz, and encoding to MP3 before embedding to Music2Latent representations. Finally, we use Deep12 \cite{deep12} to automatically extract the following meta-data: 1) segment boundaries 2) ordered segment labelings 3) beat positions 4) measures 5) chord labels.

\subsection{Model and Training}
\label{sec:modeltrain}
%model
Our model follows a standard 12-layer causal Transformer with the additions: 
For the input, we ``embed'' latent sequences with a maximum length of 4600, corresponding to approximately 7 minutes of audio, using a linear projection layer that upscales the latents' dimensions.
For the output, we use a linear layer to upscale the output dimension to the number of parameters of a 32-component GMM model. Similarly to \cite{Tshannen2023givt}, we use a diagonal covariance matrix and ensure non-negativity by using a softplus function.
%$\text{Softplus}\left(x\right) = \log\left(1+\exp\left(x\right)\right)$
We ensure the mixing coefficients are non-negative and sum to one by using a softmax function. The model is trained by minimizing \eqref{eq:ic_con}. We use Pre-Layer Normalization similar to \cite{radford2019language}, rotary positional embeddings \cite{Su2024Roformer}, and FlashAttention \cite{Dao2022Flashattention}. 

We train our model with a batch size of 16 sequences corresponding to a maximum of $\sim$ 2 hours of music. We use 16-bit automatic mixed precision and automatic gradient scaling. We use a linear learning rate schedule, which increases the learning rate from $0$ to $10^{-4}$ over 80 warmup epochs and decreases again to $0$ at epoch 500. The system is terminated with early stopping after converging at epoch 90. We use an Adam optimizer \cite{KingmaB14} with a weight decay of 3$\times 10^{-7}$. 

%For the experiment on correlation with EEG responses \cref{sec:eeg}, 
For our experiment in \cref{sec:eeg} that uses IC of singing voices, we additionally fine-tune our model on the vocal stems of our train dataset until convergence for another 69 epochs.
\section{Experiments}
\subsection{Surprisal Reduction with Repetition}
\label{sec:repetition}
\begin{table}[t]
\centering
%Anlaysis of IC on three datasets (rows), 
\caption{Repetition and contrasting segments avg. IC difference and IC/complexity correlations on three datasets (rows).}
\label{tab:res}
% \begin{tabular}{lcccccc}
% \toprule
%      & \multicolumn{1}{c}{Repetition} & \multicolumn{1}{c}{Segment} & \multicolumn{4}{c}{Complexity} \\
%  & $\rho$ & $\rho$ & $\rho_{\mathit{di}}$ & $\rho_{\mathit{rh}}$ & $\rho_{\mathit{od}}$ & $\rho_{\mathit{lo}}$  \\
%  \cmidrule(r){2-2}
%  \cmidrule(r){3-3}
%  \cmidrule(r){4-7}
% MU & -0.76 & 0.89 & 0.07 & 0.07 & -0.03 & 0.45 \\
% ME & -0.64 & 0.85 & 0.21 & 0.09 & 0.14 & 0.57 \\
% PR & -0.52 & 1 & 0.20 & 0.11 & 0.07 & 0.70 \\
% \bottomrule
% \end{tabular}
\begin{tabular}{lcccccccc}
\toprule
% Including the p-values in the table
 %     & \multicolumn{2}{c}{Repetition} & \multicolumn{4}{c}{Segment} & \multicolumn{4}{c}{Complexity} \\
 % & $\mu$ & \textit{p} & $\mu_{\mathit{s}}$ & $\mathit{p}_{\mathit{s}}$ & $\mu_{\mathit{o}}$ & $\mathit{p}_{\mathit{o}}$ & $\rho_{\mathit{d}}$ & $\rho_{\mathit{r}}$ & $\rho_{\mathit{o}}$ & $\rho_{\mathit{l}}$  \\
 % \cmidrule(r){2-3}
 % \cmidrule(r){4-7}
 % \cmidrule(r){8-11}
% MUS & -0.15 & 0.30 & 2.33 & $1.7\times 10^{-5}$ & -2.33 & $8.8\times 10^{-2}$ & 0.07 & 0.07 & -0.03 & 0.45 \\
% Med & 0.03 & 0.52 & 1.94 & $7.8\times 10^{-3}$ & -3.71 & $3.4\times 10^{-2}$ & 0.21 & 0.09 & 0.14 & 0.57 \\
% InH & -1.30 & 0 & 3.46 & 0 & -19.0 & 0 & 0.20 & 0.11 & 0.07 & 0.70 \\
% MUS & -0.15 & 0.30 & 2.33 & 1.7e-5 & -2.33 & 8.8e-2 & 0.07 & 0.07 & -0.03 & 0.45 \\
% Med & 0.03 & 0.52 & 1.94 & 7.8e-3 & -3.71 & 3.4e-2 & 0.21 & 0.09 & 0.14 & 0.57 \\
% InH & -1.30 & 0 & 3.46 & 0 & -19.0 & 0 & 0.20 & 0.11 & 0.07 & 0.70 \\
% excluding the p-values

     & \multicolumn{1}{c}{Repetition} & \multicolumn{2}{c}{Seg. Contrast} & \multicolumn{5}{c}{Complexity} \\
 & $\mu$ & $\mu_{\mathit{s}}$ & $\mu_{\mathit{o}}$ & $r_{\mathit{d}}$ & $r_{\mathit{r}}$ & $r_{\mathit{o}}$ & $r_{\mathit{l}}$ & $r_{\mathit{f}}$  \\
 \cmidrule(r){2-2}
 \cmidrule(r){3-4}
 \cmidrule(r){5-9}
MU & -0.15  & 2.33  & -2.33 & 0.07 & 0.07 & -0.03 & 0.45 & 0.38\\
ME & 0.03  & 1.94  & -3.71 & 0.21 & 0.09 & 0.14 & 0.57 & 0.56 \\
PR & -1.30 & 3.46  & -19.0 & 0.20 & 0.11 & 0.07 & 0.70 & 0.38\\
\bottomrule
\end{tabular}
\end{table}
We investigate if repetition lowers our IC as seen in symbolic music IC modeling \cite{BjareLW22, bjare2024controlling}.
In the following, a repetition is a pair of segment indices where the corresponding segments have the same label (extracted with Deep12). 
For example, a piece with musical form \textit{Intro,A,A,B,A,Outro} has two repetitions: (2,3) and (3,5), stemming from the segments labeled A. 
We quantify the IC decrease of a repetition pair by calculating the difference between its corresponding segments' average frame IC. We collect the differences for all segment types within all pieces, calculate the mean difference ($\mu$), and a one-tailed \textit{t}-test for positivity. %with the null hypothesis that the mean is 0 and the alternative hypothesis that the mean is less than 0.
The average differences are then reported in \Cref{tab:res}, column \textit{Repetition}. For PR, the decrease is negative, and the \textit{t}-test is significant; for MU and ME, the results are not significant. In the following, we use 5\% significance levels unless otherwise stated. Therefore, repetition lowers IC on average for the largest dataset, which follows the training distribution most closely. We emphasize that our results are obtained for repetitions that do not correspond to exact copies since the data is recorded music.  

\subsection{Contrasting Segments Hypothesis}
We are interested in the following hypothesis: segment types, as identified by their labels, contrast their predecessor segment types in the sense of having a higher IC on average. Outros, however, serve to close a piece and, therefore, have lower IC than predecessor segments. As for repetitions, our experiments involve calculating the difference in segment ICs. However, instead of extracting repetition index pairs, we extract index pairs corresponding to subsequent sequence types on their first occurrence.
For the previous example, we, therefore, obtain index pairs (1,2), (2,4) and (4,6).
We calculate an average across all pieces and report this in \Cref{tab:res}, column Seg. Contrast by: $\mu_{\mathit{s}}$ considering all segment types not including outros, and $\mu_{\mathit{o}}$ only segment pairs that involve outros. For segment types not involving outros, we find that $\mu_{\mathit{s}}$ is positive for all datasets, and the results are significant using a one-tailed $t$-test. For segments involving outros, we find all $\mu_{\mathit{o}}$ to be negative and significant except for MU. The results, therefore, support our hypothesis.

\subsection{Relation to Complexity}
\label{sec:correlation}
\begin{figure}
    \centering
    \includegraphics[trim=1.0cm 0.0 .7cm 1.2cm, clip, width=\linewidth]{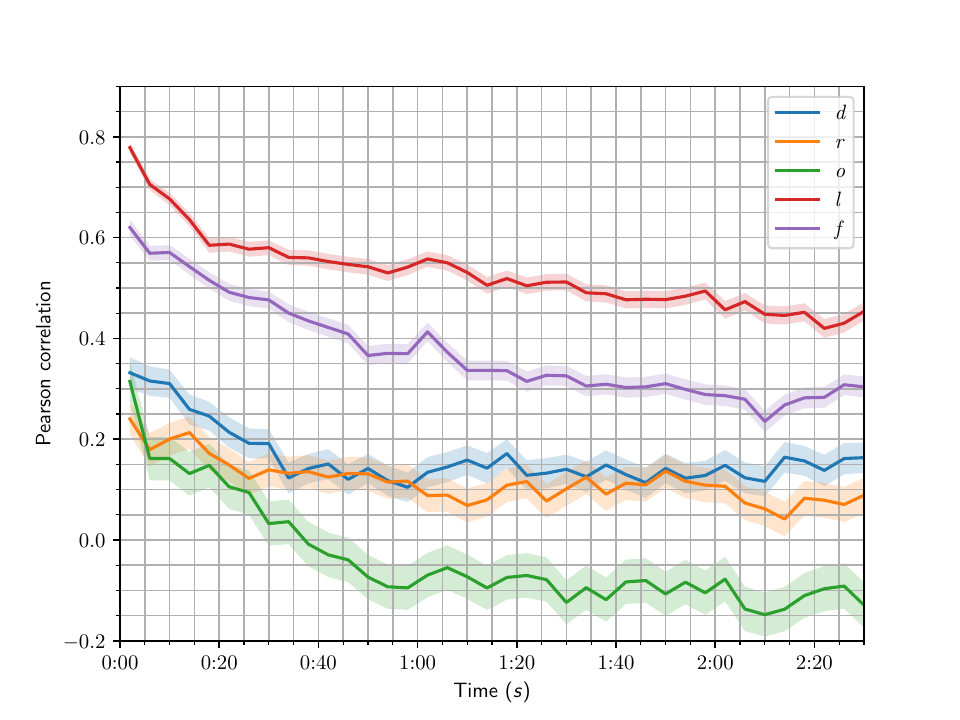}
    % \caption{Temporal development of the complexity correlations on PR dataset.}
    \caption{Temporal development of correlations between IC and dissonance $(d)$, IOI-entropy $(r)$, onset density $(o)$ loudness $(l)$ and spectral flux $(f)$ on the dataset PR. %\stefan{It's best if the caption is as self-explanatory as possible. What's d,l,o,r,f and better write ``on the PR dataset'' and correlations between what? Same for Table I}
    }
    \label{fig:lc}
\end{figure}
%\mb{Is complexity the right term to use? For instance, are loudness and timbral variance a complexity? Does anyone have a suggestion for a broader term?}
We are interested in the relationship between the IC and descriptive features of the music, which can be associated with complexity. As such, we investigated to what extent IC is correlated with dissonance ($d$), as indicated by TIV features \cite{bernardes2015conchord,bernardes2016multi}, rhythmic complexity ($r$) as indicated by the normalized entropy of interonset interval (IOI) histogram \cite{ioient}, onset density ($o$) and spectral flux ($f$) associated with timbre variations \cite{mcadams1995perceptual}. Additionally, we investigate IC's relation to the signal's loudness ($l$). % We do so by extracting windows of metric-dependent lengths, calculating the metric and mean IC of the segment's frames, and reporting Pearson's correlation. 
We do so by extracting windows of metric-dependent lengths and calculating Pearson's correlation between the metric and the mean IC of the segment's frames. 
% This is done for each dataset. 
We calculated the TIV dissonance of 2-second windows centered around the chord changes within the pieces, because we hypothesized that the greatest dissonance variations would be found there. The IOI-entropies are calculated by quantizing onsets calculated with the detection algorithm of \cite{bock2013maximum} to 12 subdivisions per beat and calculating the normalized entropy of IOI histograms within each measure. The onset densities are calculated using 4-second windows. Loudness is calculated on 1-second windows according to \cite{bs1770}.
%Finally, we calculate the 
Average spectral flux is calculated using 1-second bin-normalized log-mel spectrogram windows.
The results are presented in \Cref{tab:res}, column Complexity, where $r_{\mathit{d}},r_{\mathit{r}}, r_{\mathit{o}}, r_{\mathit{l}}$ and $r_{\mathit{f}}$ are the correlations corresponding to TIV dissonance, IOI-entropy, onset density, loudness and spectral flux respectively. All correlations are significant. 
Loudness and spectral flux are mostly correlated with the IC across all datasets. For loudness, possibly due to quieter parts corresponding to segments with fewer simultaneously sounding stems with less information than louder segments. For spectral flux, it indicates that timbre variations influence surprisal. Dissonance is the third highest correlated metric, with a correlation of 0.2 on ME and PR. The rest of the correlations are neglectable. As shown in \cref{sec:repetition}, repetition decreases the IC values. We, therefore, investigate how the correlations between IC and the metrics develop over time. To that end, we bin the timeline in 4-second intervals and associate the previously mentioned windows with a particular bin if its start point is within that window. For each window bin, we calculated the correlations for the dataset PR and report it in \Cref{fig:lc}. We present the results for PR since it contains most pieces, but the other datasets showed similar trends. The correlations are noticeably higher at the beginning of pieces than the average correlations of \Cref{tab:res} for all metrics; however, the correlations decrease over time. This is consistent with IC complexity correlations in the symbolic domain \cite{bjare2024controlling}. Since the data is popular music, it is reasonable to believe that many pieces' ICs are dominated by repetitions after the first minute, which decreases the IC to some lower bound, resulting in a decrease in the correlation. 
% \mb{As such, our IC, like the symbolic analogous, is sensitive to the complexity of the musical content; however, it is additionally sensitive to aspects of the recording that cannot be represented in symbolic music, thus opening new dimensions to the study of computational models of surprisal.}

\subsection{Prediction of EEG Responses}
\label{sec:eeg}
To validate whether the proposed modeling correlates with human perception, we test if the IC can predict neural responses to monophonic sung music.
To do so, 64-channel EEG responses to songs ($20$ adult participants, $18$ different song stimuli; for details, see \cite{cantisani2023investigating}) are modeled as a linear combination of two predictor variables, the IC and the energy envelope of the waveform (E), so that these two compete to explain EEG variance. E is extracted using the Hilbert transform and used as a nuisance regressor, absorbing EEG variance relating to lower-level acoustic features and enabling us to isolate the encoding of the higher-level processes related to expectations. Brain responses are sensitive to onsets and variations, and since the IC is correlated with loudness (see \cref{sec:correlation}), controlling for acoustic changes is crucial. To further control that, we do a constant-value interpolation of unvoiced segments using the last IC value before the silence.

The channel-specific mapping between predictors and the neural data is estimated for each participant by solving a regularised linear regression problem \cite{crosse2016multivariate}. The model accounts for non-instantaneous interactions as it is learned considering multiple stimulus-response time lags. We set a $[-100, 700]$ms time window with a margin of $50$ms to avoid border artifacts. The resulting forward models are then evaluated in terms of how well they predict unseen EEG data using a leave-one-out cross-validation procedure across trials. The quality of a prediction is then quantified by calculating the z-transformed Pearson’s correlation ($r$) between the estimated signals and the corresponding predictions at each scalp electrode. 

The significance of the IC contribution was assessed by comparing the predictive power of the full model to the average of $N=100$ baseline models consistent with the null hypothesis that there is no causal relationship between IC and brain response. The baseline models are identical to the full model except for the IC, which is randomly shuffled. %Thus, IC is considered encoded in the EEG if it significantly improves the model's fit across participants. 
This procedure enabled us to evaluate any incremental improvement in the model's performance due to a significant relationship between IC and responses without introducing bias (e.g., by altering the model's complexity) and has been widely employed in cortical tracking experiments, among which those testing IDyOM \cite{di2020cortical}. Statistical analyses use two-tailed \textit{t}-tests or non-parametric Wilcoxon signed-rank tests and corresponding effect size measures (Cohen’s \textit{d} and Wilcoxon’s Z score respectively) depending on the normality of the data, assessed via the Anderson-Darling test. False discovery rate (FDR) is used to correct for multiple comparisons.

\begin{figure}
    \centering
    \includegraphics[trim=0.2cm 0.0 .2cm 0cm, clip, width=0.7\linewidth]{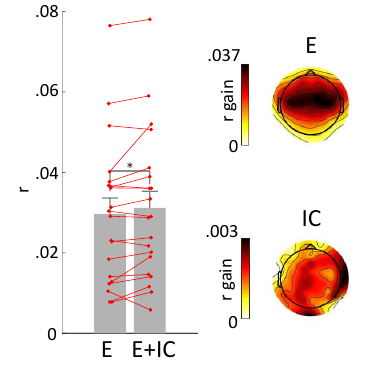}
    \caption{z-transformed Pearson's correlation gain for individual channels on scalp topographies (FDR corrected \textit{t}-test with significance threshold set at $p<0.05$) and for the average over all channels (bar plot, mean+-SE across participants and trials, red dots represent individual participants).}
    \label{fig:eeg}
\end{figure}

As can be seen in Fig.~\ref{fig:eeg}, including IC among the predictors significantly improves the average model predictions across channels (Wilcoxon signed-rank test, $p=0.025$, $Z=2.22$) and almost all individual channel predictions (FDR corrected \textit{t}-test with significance threshold set at $p=0.05$), meaning IC explains a distinct portion of the variance in the data. Note that the encoding of the control variable E is already significant \textit{per se} on the average over all electrodes (\textit{t}-test, $p<0.001$, $d=1.41$). %Finally, IC better reconstructs the right channels as opposed to simple acoustics. Yet, we found no significant effect of right-lateralization across subjects. 
E and IC topographies are consistent with those of \cite{cantisani2024neural}, where right lateralization was observed in the same data for melodic expectations computed with IDyOM and not for lower-level acoustics. Here, however, we found no significant effect of right-lateralization across subjects using the same test. 

%Here, we use the same test of hemispheric asymmetry, which compares the electrode-wise ﬁt improvement between the two hemispheres with a two-tailed \textit{t}-test while controlling for multiple comparisons. 

\section{Conclusion}
We presented a general model for estimating musical surprisal in full-length music via the IC of a GIVT model's next-step predictions on invertible audio representations. This is opposed to models of previous works that operate on preselected audio features capable of capturing only selected aspects of the audio signal. We showed that the model's IC, like symbolic music analogous models, decreases with repetition. We showed that the IC of segment types appearing later in a piece is higher than those appearing earlier on average but that the opposite is true for outros. We showed correlations between the IC and descriptive features of the music and found these to decrease with time, similar to findings in the symbolic domain. Finally, we validated our model's capabilities as a perceptual model of human musical surprisal by showing that the IC significantly predicts EEG brain responses to songs.

\clearpage

% \label{sec:refs}

\bibliographystyle{IEEEbib}
\bibliography{strings,refs}

\begin{thebibliography}{10}
\providecommand{\url}[1]{#1}
\csname url@samestyle\endcsname
\providecommand{\newblock}{\relax}
\providecommand{\bibinfo}[2]{#2}
\providecommand{\BIBentrySTDinterwordspacing}{\spaceskip=0pt\relax}
\providecommand{\BIBentryALTinterwordstretchfactor}{4}
\providecommand{\BIBentryALTinterwordspacing}{\spaceskip=\fontdimen2\font plus
\BIBentryALTinterwordstretchfactor\fontdimen3\font minus \fontdimen4\font\relax}
\providecommand{\BIBforeignlanguage}[2]{{%
\expandafter\ifx\csname l@#1\endcsname\relax
\typeout{** WARNING: IEEEtran.bst: No hyphenation pattern has been}%
\typeout{** loaded for the language `#1'. Using the pattern for}%
\typeout{** the default language instead.}%
\else
\language=\csname l@#1\endcsname
\fi
#2}}
\providecommand{\BIBdecl}{\relax}
\BIBdecl

\bibitem{meyer}
L.~B. Meyer, ``Meaning in music and information theory,'' \emph{The Journal of Aesthetics and Art Criticism}, vol.~15, no.~4, pp. 412--424, 1957.

\bibitem{idyom_conklin}
D.~Conklin and I.~H. Witten, ``Multiple viewpoint systems for music prediction,'' \emph{Journal of New Music Research}, vol.~24, no.~1, pp. 51--73, 1995.

\bibitem{idyom}
M.~Pearce, ``The construction and evaluation of statistical models of melodic structure in music perception and composition,'' Ph.D. dissertation, Department of Computing, City University, London, UK, 2005.

\bibitem{musexp}
M.~T. Pearce and G.~A. Wiggins, ``Auditory expectation: The information dynamics of music perception and cognition,'' \emph{Top. Cogn. Sci.}, vol.~4, no.~4, pp. 625--652, 2012.

\bibitem{complic}
S.~A. Sauv{\'e} and M.~T. Pearce, ``Information-theoretic modeling of perceived musical complexity,'' \emph{Music Perception: An Interdisciplinary Journal}, vol.~37, no.~2, pp. 165--178, 2019.

\bibitem{BjareLW23}
M.~R. Bjare, S.~Lattner, and G.~Widmer, ``Exploring sampling techniques for generating melodies with a transformer language model,'' in \emph{{ISMIR}}, 2023, pp. 810--816.

\bibitem{gold2019predictability}
B.~P. Gold, M.~T. Pearce, E.~Mas-Herrero, A.~Dagher, and R.~J. Zatorre, ``Predictability and uncertainty in the pleasure of music: a reward for learning?'' \emph{Journal of Neuroscience}, vol.~39, no.~47, pp. 9397--9409, 2019.

\bibitem{wang2014guided}
C.-i. Wang and S.~Dubnov, ``Guided music synthesis with variable markov oracle,'' in \emph{{AAAI}}, vol.~10, no.~5, 2014, pp. 55--62.

\bibitem{collins2016developing}
T.~Collins, R.~Laney, A.~Willis, and P.~H. Garthwaite, ``Developing and evaluating computational models of musical style,'' \emph{AI EDAM}, vol.~30, no.~1, pp. 16--43, 2016.

\bibitem{bjare2024controlling}
M.~R. Bjare, S.~Lattner, and G.~Widmer, ``Controlling surprisal in music generation via information content curve matching,'' in \emph{{ISMIR}}, 2024.

\bibitem{dubnov2007audio}
S.~Dubnov, G.~Assayag, and A.~Cont, ``Audio oracle: A new algorithm for fast learning of audio structures,'' in \emph{{ICMC}}, 2007, pp. 224--227.

\bibitem{masclef2023deep}
N.~L. Masclef and T.~A. Keller, ``Deep generative models of music expectation,'' \emph{NeurIPS ML for Audio Workshop 2023}, 2023.

\bibitem{abrams2022retrieving}
E.~Abrams, E.~M. Vidal, C.~Pelofi, and P.~Ripoll{\'e}s, ``Retrieving musical information from neural data: how cognitive features enrich acoustic ones.'' in \emph{{ISMIR}}, 2022, pp. 160--168.

\bibitem{BjareLW22}
M.~R. Bjare, S.~Lattner, and G.~Widmer, ``Differentiable short-term models for efficient online learning and prediction in monophonic music,'' \emph{Trans. Int. Soc. Music. Inf. Retr.}, vol.~5, no.~1, p. 190, 2022.

\bibitem{bjarelattner}
M.~R. Bjare and S.~Lattner, ``On the typicality of musical sequences,'' in \emph{ISMIR Late Breaking and Demo Papers}, 2022.

\bibitem{VaswaniSPUJGKP17}
A.~Vaswani, N.~Shazeer, N.~Parmar, J.~Uszkoreit, L.~Jones, A.~N. Gomez, L.~Kaiser, and I.~Polosukhin, ``Attention is all you need,'' in \emph{NeurIPS}, 2017, pp. 5998--6008.

\bibitem{evans2024long}
Z.~Evans, J.~D. Parker, C.~Carr, Z.~Zukowski, J.~Taylor, and J.~Pons, ``Long-form music generation with latent diffusion,'' in \emph{{ISMIR}}, 2024.

\bibitem{Tshannen2023givt}
M.~Tschannen, C.~Eastwood, and F.~Mentzer, ``{GIVT:} generative infinite-vocabulary transformers,'' \emph{CoRR}, vol. abs/2312.02116, 2023.

\bibitem{pasini2024music2latent}
M.~Pasini, S.~Lattner, and G.~Fazekas, ``Music2latent: Consistency autoencoders for latent audio compression,'' in \emph{{ISMIR}}, 2024.

\bibitem{pitch_per}
M.~T. Pearce, M.~H. Ruiz, S.~Kapasi, G.~A. Wiggins, and J.~Bhattacharya, ``Unsupervised statistical learning underpins computational, behavioural, and neural manifestations of musical expectation,'' \emph{NeuroImage}, vol.~50, no.~1, pp. 302--313, 2010.

\bibitem{di2020cortical}
G.~M. Di~Liberto, C.~Pelofi, R.~Bianco, P.~Patel, A.~D. Mehta, J.~L. Herrero, A.~De~Cheveign{\'e}, S.~Shamma, and N.~Mesgarani, ``Cortical encoding of melodic expectations in human temporal cortex,'' \emph{Elife}, vol.~9, p. e51784, 2020.

\bibitem{hansen2014predictive}
N.~C. Hansen and M.~T. Pearce, ``Predictive uncertainty in auditory sequence processing,'' \emph{Frontiers in psychology}, vol.~5, p. 1052, 2014.

\bibitem{bianco2020pupil}
R.~Bianco, L.~E. Ptasczynski, and D.~Omigie, ``Pupil responses to pitch deviants reflect predictability of melodic sequences,'' \emph{Brain and Cognition}, vol. 138, p. 103621, 2020.

\bibitem{moldwin2017statistical}
T.~Moldwin, O.~Schwartz, and E.~S. Sussman, ``Statistical learning of melodic patterns influences the brain's response to wrong notes,'' \emph{Journal of cognitive neuroscience}, vol.~29, no.~12, pp. 2114--2122, 2017.

\bibitem{Skerritt-DavisE18}
B.~Skerritt{-}Davis and M.~Elhilali, ``Detecting change in stochastic sound sequences,'' \emph{PLoS Comput. Biol.}, vol.~14, no.~5, 2018.

\bibitem{skerritt2019model}
B.~Skerritt-Davis and M.~Elhilali, ``A model for statistical regularity extraction from dynamic sounds,'' \emph{Acta Acustica united with Acustica}, vol. 105, no.~1, pp. 1--4, 2019.

\bibitem{SongD0S23}
Y.~Song, P.~Dhariwal, M.~Chen, and I.~Sutskever, ``Consistency models,'' in \emph{{ICML}}, vol. 202.\hskip 1em plus 0.5em minus 0.4em\relax {PMLR}, 2023, pp. 32\,211--32\,252.

\bibitem{BaevskiZMA20}
A.~Baevski, Y.~Zhou, A.~Mohamed, and M.~Auli, ``wav2vec 2.0: {A} framework for self-supervised learning of speech representations,'' in \emph{NeurIPS}, 2020, pp. 12\,449--12\,460.

\bibitem{SpijkervetB21}
J.~Spijkervet and J.~A. Burgoyne, ``Contrastive learning of musical representations,'' in \emph{{ISMIR}}, 2021, pp. 673--681.

\bibitem{jukebox}
P.~Dhariwal, H.~Jun, C.~Payne, J.~W. Kim, A.~Radford, and I.~Sutskever, ``Jukebox: {A} generative model for music,'' \emph{CoRR}, vol. abs/2005.00341, 2020.

\bibitem{ZeghidourLOST22}
N.~Zeghidour, A.~Luebs, A.~Omran, J.~Skoglund, and M.~Tagliasacchi, ``Sound{S}tream: An end-to-end neural audio codec,'' \emph{{IEEE} {ACM} Trans. Audio Speech Lang. Process.}, vol.~30, pp. 495--507, 2022.

\bibitem{DBLP:journals/tmlr/DefossezCSA23}
A.~D{\'{e}}fossez, J.~Copet, G.~Synnaeve, and Y.~Adi, ``High fidelity neural audio compression,'' \emph{Trans. Mach. Learn. Res.}, vol. 2023, 2023.

\bibitem{KumarSLKK23}
R.~Kumar, P.~Seetharaman, A.~Luebs, I.~Kumar, and K.~Kumar, ``High-fidelity audio compression with improved {RVQGAN},'' in \emph{NeurIPS}, 2023.

\bibitem{radford2019language}
A.~Radford, J.~Wu, R.~Child, D.~Luan, D.~Amodei, and I.~Sutskever, ``Language models are unsupervised multitask learners,'' 2019.

\bibitem{bittner2016medleydb}
R.~M. Bittner, J.~Wilkins, H.~Yip, and J.~P. Bello, ``Medleydb 2.0: New data and a system for sustainable data collection,'' \emph{ISMIR Late Breaking and Demo Papers}, vol.~36, 2016.

\bibitem{musdb18}
\BIBentryALTinterwordspacing
Z.~Rafii, A.~Liutkus, F.-R. St{\"o}ter, S.~I. Mimilakis, and R.~Bittner, ``The {MUSDB18} corpus for music separation,'' Dec. 2017. [Online]. Available: \url{https://doi.org/10.5281/zenodo.1117372}
\BIBentrySTDinterwordspacing

\bibitem{deep12}
H.~Kishi, N.~Polouliakh, and T.~Akama, ``Deep12,'' https://www.sonycsl.co.jp/tokyo/14621.

\bibitem{Su2024Roformer}
J.~Su, M.~H.~M. Ahmed, Y.~Lu, S.~Pan, W.~Bo, and Y.~Liu, ``Roformer: Enhanced transformer with rotary position embedding,'' \emph{Neurocomputing}, vol. 568, p. 127063, 2024.

\bibitem{Dao2022Flashattention}
T.~Dao, D.~Y. Fu, S.~Ermon, A.~Rudra, and C.~R{\'{e}}, ``Flash{A}ttention: Fast and memory-efficient exact attention with io-awareness,'' \emph{CoRR}, vol. abs/2205.14135, 2022.

\bibitem{KingmaB14}
D.~P. Kingma and J.~Ba, ``Adam: {A} method for stochastic optimization,'' in \emph{{ICLR} (Poster)}, 2015.

\bibitem{bernardes2015conchord}
G.~Bernardes, D.~Cocharro, C.~Guedes, and M.~Davies, ``Conchord: an application for generating musical harmony by navigating in a perceptually motivated tonal interval space,'' in \emph{{CMMR}}, 2015, pp. 71--86.

\bibitem{bernardes2016multi}
G.~Bernardes, D.~Cocharro, M.~Caetano, C.~Guedes, and M.~E. Davies, ``A multi-level tonal interval space for modelling pitch relatedness and musical consonance,'' \emph{Journal of New Music Research}, vol.~45, no.~4, pp. 281--294, 2016.

\bibitem{ioient}
A.~A. Moles, \emph{Information theory and esthetic perception}.\hskip 1em plus 0.5em minus 0.4em\relax The University of Illinois Press, Urbana and London, 1966.

\bibitem{mcadams1995perceptual}
S.~McAdams, S.~Winsberg, S.~Donnadieu, G.~De~Soete, and J.~Krimphoff, ``Perceptual scaling of synthesized musical timbres: Common dimensions, specificities, and latent subject classes,'' \emph{Psychological research}, vol.~58, pp. 177--192, 1995.

\bibitem{bock2013maximum}
S.~B{\"o}ck and G.~Widmer, ``Maximum filter vibrato suppression for onset detection,'' in \emph{DAFx}, vol.~7.\hskip 1em plus 0.5em minus 0.4em\relax Citeseer, 2013, p.~4.

\bibitem{bs1770}
{ITU Radiocommunication Sector - Broadcasting service (sound)}, ``{ITU-R BS.1770-4},'' https://www.itu.int/rec/R-REC-BS.1770/en.

\bibitem{cantisani2023investigating}
G.~Cantisani, A.~Chalehchaleh, G.~Di~Liberto, and S.~Shamma, ``Investigating the cortical tracking of speech and music with sung speech,'' in \emph{{INTERSPEECH}}.\hskip 1em plus 0.5em minus 0.4em\relax {ISCA}, 2023, pp. 5157--5161.

\bibitem{crosse2016multivariate}
M.~J. Crosse, G.~M. Di~Liberto, A.~Bednar, and E.~C. Lalor, ``The multivariate temporal response function (mtrf) toolbox: a matlab toolbox for relating neural signals to continuous stimuli,'' \emph{Frontiers in human neuroscience}, vol.~10, p. 604, 2016.

\bibitem{cantisani2024neural}
G.~Cantisani, S.~Shamma, and G.~M. Di~Liberto, ``Neural signatures of musical and linguistic interactions during natural song listening,'' \emph{Hal preprint}, 2024.

\end{thebibliography}

\end{document}